\documentclass[reprint,superscriptaddress,longbibliography,amsmath,amssymb]{revtex4-2}

\usepackage{graphicx}
\usepackage{siunitx}
\usepackage{booktabs}
\usepackage{textcomp}
\usepackage{microtype}
\usepackage[colorlinks=true,linkcolor=blue,citecolor=blue,urlcolor=blue]{hyperref}

\DeclareSIUnit\HV{HV}
\DeclareSIUnit\kgf{kgf}
\DeclareSIUnit\bar{bar}
\DeclareSIUnit\Angstrom{\text{\AA}}
\DeclareSIUnit\muB{\ensuremath{\mu_\mathrm{B}}}
\providecommand{\micro}{\textmu}

\begin{document}

\title{Probing the transport properties of Cantor--Wu alloys by means of femtosecond and nanosecond laser ablation}

\author{David Redka}
\email{dredka@hm.edu}
\affiliation{Laser Center HM, Munich University of Applied Sciences HM, Lothstra\ss e 34, 80335 Munich, Germany}

\author{Maximilian Spellauge}
\affiliation{Laser Center HM, Munich University of Applied Sciences HM, Lothstra\ss e 34, 80335 Munich, Germany}

\author{Rosemary Babu}
\affiliation{Laser Center HM, Munich University of Applied Sciences HM, Lothstra\ss e 34, 80335 Munich, Germany}

\author{Christopher D. Woodgate}
\affiliation{H. H. Wills Physics Laboratory, University of Bristol, Tyndall Avenue, BS8 1TL Bristol, United Kingdom}

\author{Hubert Ebert}
\affiliation{Department of Chemistry, Ludwig-Maximilians-Universit\"at M\"unchen, Butenandtstra\ss e 5--13, 81377 Munich, Germany}

\author{J\'an Min\'ar}
\affiliation{New Technologies Research Centre, University of West Bohemia, Univerzitn\'i 8, 301 00 Pilsen, Czech Republic}

\author{Daniel J. F\"orster}
\affiliation{Laser Center HM, Munich University of Applied Sciences HM, Lothstra\ss e 34, 80335 Munich, Germany}

\author{Heinz P. Huber}
\affiliation{Laser Center HM, Munich University of Applied Sciences HM, Lothstra\ss e 34, 80335 Munich, Germany}
\affiliation{New Technologies Research Centre, University of West Bohemia, Univerzitn\'i 8, 301 00 Pilsen, Czech Republic}

\date{\today}

\begin{abstract}
Single-pulse laser ablation thresholds of selected equiatomic Cantor--Wu alloys---FeNi, CoNi, CrFeNi, CrCoNi, and CrMnFeCoNi---are measured for femtosecond and nanosecond pulse durations and interpreted through first-principles calculations of the electronic structure, the electron--phonon coupling, and the electronic thermal conductivity. Alloy synthesis, ablation experiments, and theory are performed consistently on the same set of samples. The absorbed femtosecond thresholds decrease systematically by up to \SI{36}{\percent} from FeNi to the Cr-containing alloys, a trend that reflectance variations cannot explain. Two-temperature-model scaling of the thresholds with the electronic thermal conductivity and the electron--phonon coupling, with all parameters taken from the spin-disordered phase, reproduces the measured hierarchy. The nanosecond thresholds instead probe the thermal equilibrium conductivity averaged along the heating path. The apparent outlier of CoNi, whose room-temperature transport over-predicts its thresholds by up to a factor of two for both pulse durations, is resolved quantitatively by the collapse of its conductivity upon loss of ferromagnetic order. Single-pulse ablation thresholds thereby emerge as sensitive, contact-free probes of electronic transport and of its magnetic-phase dependence in compositionally complex alloys.
\end{abstract}


\maketitle

\section{Introduction}
Since the concurrent reports of Cantor \textit{et al.}~\cite{cantorMicrostructuralDevelopmentEquiatomic2004} and Yeh \textit{et al.}~\cite{yehNanostructuredHighEntropyAlloys2004}, the study of multi-principal-element and high-entropy alloys has developed into one of the most active fields of metals research~\cite{miracleCriticalReviewHigh2017,georgeHighentropyAlloys2019}. What makes these materials attractive is that a single crystal structure can host a vast range of compositions, and that some of these compositions combine high yield strength with an unusual resistance to fracture down to cryogenic temperatures, most prominently CrMnFeCoNi and CrCoNi~\cite{gludovatzFractureresistantHighentropyAlloy2014,gludovatzExceptionalDamagetoleranceMediumentropy2016,ottoInfluencesTemperatureMicrostructure2013}. Within this compositional space, the equiatomic Cantor--Wu alloys, the face-centred cubic single-phase family spanning the binary alloys FeNi and CoNi, the ternary alloys CrFeNi and CrCoNi, and the quinary CrMnFeCoNi~\cite{wuTemperatureDependenceMechanical2014,wuRecoveryRecrystallizationGrain2014,laplancheElasticModuliThermal2018}, have become the model system of choice for isolating the influence of chemical complexity at fixed crystal structure. Their functional properties respond far more strongly to composition than their shared crystal structure suggests. The Cr-containing alloys show the highest yield strength of the series~\cite{wuTemperatureDependenceMechanical2014}, while ductility and work-hardening capability are retained~\cite{gludovatzFractureresistantHighentropyAlloy2014}. The magnetic properties change just as strongly, from ferromagnetic order in FeNi and CoNi with Curie temperatures far above room temperature to paramagnetic behaviour of the Cr-containing alloys, whose magnetic ordering temperatures drop to a few tens of Kelvin~\cite{jinThermophysicalPropertiesNicontaining2017,schneeweissMagneticPropertiesCrMnFeCoNi2017, billingtoBulkElementSpecificMagnetism2020}. The electronic transport follows the same pattern. Adding Cr and Mn turns the weakly scattering ferromagnetic metals into strongly disordered paramagnets, changing the electrical resistivity and the electronic thermal conductivity by up to an order of magnitude~\cite{jinTailoringPhysicalProperties2016,muUncoveringElectronScattering2019,samolyukTemperatureDependentElectronic2018,redkaInterplayDisorderElectronic2024}.

Laser processing is a natural machining route for these hard and ductile alloys. Ultrashort-pulse ablation of high-entropy alloys and related 3$d$ transition-metal alloys is receiving growing attention~\cite{redkaSubpicosecondSinglepulseLaser2021,chenTimeresolvedProbingModeling2025,thomaeDecipheringDrivingMechanisms2026}, and the single-pulse ablation threshold fluence is the key parameter of any such process. For ultrashort pulses, the dynamics proceed under strong electron--lattice nonequilibrium, described by the two-temperature model (TTM), in which the electronic thermal conductivity $\kappa_\mathrm{e}$ and the electron--phonon coupling factor $G_\mathrm{ep}$ set the competition between diffusive energy dissipation into the bulk and energy transfer to the lattice, respectively~\cite{chichkovFemtosecondPicosecondNanosecond1996,nolteAblationMetalsUltrashort1997,wellershoffRoleElectronPhonon1999,hohlfeldElectronLatticeDynamics2000,rethfeldModellingUltrafastLaser2017}. Material removal near threshold then proceeds by spallation, that is, the ejection of a near-surface liquid layer upon tensile unloading of the laser-induced GPa-level stresses~\cite{inogamovNanospallationInducedUltrashort2008,zhigileiAtomisticModelingShort2009,linUltrafastMeltingSpallation2025}, and, at higher fluences, by phase explosion of overheated matter~\cite{ohFemtosecondLaserAblation2007,winterUltrashortSinglepulseLaser2021}. For ultrashort pulses, threshold differences between metals are therefore known to track these transport parameters~\cite{wellershoffRoleElectronPhonon1999,winterUltrashortSinglepulseLaser2021}. For nanosecond pulses, by contrast, electrons and lattice remain essentially in equilibrium, ablation is governed by quasi-equilibrium heating, melt formation, and melt expulsion, and the threshold instead probes the equilibrium thermal conductivity~\cite{corkumThermalResponseMetals1988,stuartNanosecondtofemtosecondLaserinducedBreakdown1996,mommaShortpulseLaserAblation1996}. For the Cantor--Wu series of binary, medium-entropy, and high-entropy alloys, however, only isolated ultrashort-pulse thresholds of individual alloys have been reported~\cite{redkaSubpicosecondSinglepulseLaser2021,chenTimeresolvedProbingModeling2025}, a systematic single-pulse study across the full series is lacking, and nanosecond thresholds---probing an entirely different ablation mechanism and dynamics---are missing altogether.

From a theoretical perspective, the electronic structure and transport properties of compositionally complex alloys are accessible from first principles, but the number of possible atomic arrangements makes supercell calculations that average over random configurations computationally demanding~\cite{muUncoveringElectronScattering2019}. An efficient and well-established approach to treat chemical disorder is provided by the coherent potential approximation (CPA)~\cite{sovenCoherentPotentialModelSubstitutional1967,faulknerCalculatingPropertiesCoherentpotential1980,ebertRelativisticBandstructureDisordered1997}, as implemented within the relativistic Korringa--Kohn--Rostoker (KKR) Green's-function method~\cite{ebertCalculatingCondensedMatter2011}. Spin disorder is modelled via the disordered-local-moment (DLM) picture~\cite{pindorDisorderedLocalMoment1983,gyorffyFirstprinciplesTheoryFerromagnetic1985}, while thermal lattice and spin disorder are described within the alloy-analogy model~\cite{ebertCalculatingLinearresponseFunctions2015}. Transport properties follow from linear-response theory~\cite{butlerTheoryElectronicTransport1985}. For the Cantor--Wu alloys, such calculations reproduce the measured resistivity trends and have revealed the decisive role of magnetic order, with the binary alloys owing their high conductivity to a weakly scattering majority-spin channel, while Cr additions disorder both spin channels and smear the Fermi surface~\cite{muUncoveringElectronScattering2019,samolyukTemperatureDependentElectronic2018,jinTailoringPhysicalProperties2016,robartsExtremeFermiSurface2020,redkaInterplayDisorderElectronic2024}. In parallel, first-principles descriptions of the electronic heat capacity~\cite{bevillonFirstprinciplesCalculationsHeat2015} and of the electron--phonon coupling~\cite{linElectronphononCouplingElectron2008} for elemental metals have been extended to alloys such as stainless steels~\cite{winterTemperaturedependentDeterminationElectron2016}.

Whether measured ablation thresholds of a chemically complex alloy family actually follow these calculated transport parameters has, however, never been tested directly, and for the Cantor--Wu series this raises a particularly pointed question. The absorption is nearly composition independent, while the transport parameters vary by an order of magnitude and depend sensitively on the magnetic phase, which femtosecond excitation itself destroys within a few hundred femtoseconds through ultrafast demagnetization~\cite{beaurepaireUltrafastSpinDynamics1996a,koopmansExplainingParadoxicalDiversity2010,eichBandStructureEvolution2017}. Which magnetic phase sets the transport parameters that govern the ablation threshold, and whether thresholds can conversely serve as quantitative probes of transport in compositionally complex alloys, has remained open.

Here, femtosecond and nanosecond single-pulse ablation threshold measurements of five selected Cantor--Wu alloys are combined with first-principles calculations of the electronic structure, the electron--phonon coupling, the electronic and optical conductivity, and the magnetization to address the issues and questions outlined above. All steps are performed consistently within this work on the same set of samples, from alloy synthesis and characterization through the ablation experiments for both pulse durations to the first-principles theory. Single-pulse ablation thresholds thereby emerge as sensitive, contact-free probes of electronic transport and of its magnetic-phase dependence in compositionally complex alloys.

\section{Methods}
\subsection{Sample preparation and characterization}
Master alloys (\SI{35}{\gram} total mass) were cast from high-purity (\SI{99.99}{\percent}) Cr, Mn, Fe, Co, and Ni pellets, weighed to $\pm\SI{0.1}{\milli\gram}$ on a precision balance. Arc melting was carried out on a water-cooled copper hearth in a Bühler furnace. The chamber was evacuated to $<\SI{0.1}{\milli\pascal}$, backfilled with high-purity Ar, and stabilized at \SI{500}{\milli\bar}, after which a DC arc at \SI{150}{\ampere} and \SI{12}{\volt} was struck. Each button was fully melted, mechanically flipped, and remelted twice to promote chemical homogeneity and suppress segregation. The as-cast buttons were then homogenized at \SI{1500}{\kelvin} for \SI{48}{\hour} under a vacuum of \SI{0.1}{\milli\pascal}. Disc-shaped specimens (\SI{1.0}{\milli\metre} thickness, \SI{20}{\milli\metre} diameter) were sectioned by wire electrical discharge machining, embedded in epoxy resin, ground with SiC paper (grit 400 and 1000), and polished sequentially with polycrystalline diamond suspensions (\SI{9}{\micro\metre}, \SI{3}{\micro\metre}, \SI{1}{\micro\metre}) to a mirror finish.

\begin{figure}[!ht]
  \centering
  \includegraphics[width=\linewidth]{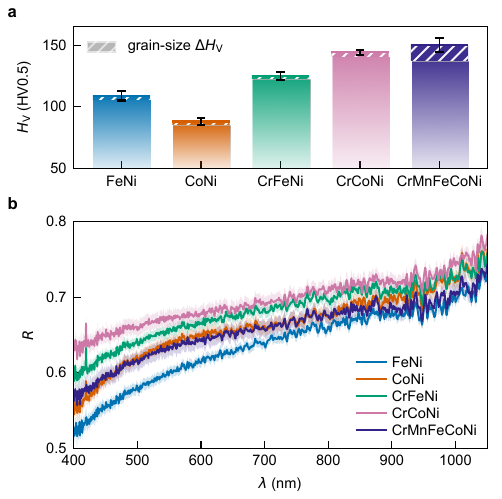}
  \caption{Surface properties of the polished selected Cantor--Wu alloys (FeNi, CoNi, CrFeNi, CrCoNi, CrMnFeCoNi). \textbf{a,} Vickers microhardness $H_\mathrm{V}$ (load HV0.5, mean of five indents per alloy, error bars the standard deviation). The solid bar is the intrinsic hardness, $H_\mathrm{V0}$, and the hatched cap the grain-size (Hall--Petch) contribution $\Delta H_\mathrm{V}=k_\mathrm{H}\,d_\mathrm{gr}^{-1/2}$ obtained from the measured grain size, $d_\mathrm{gr}$. \textbf{b,} Normal-incidence reflectance, $R$, of the same surfaces over \SIrange{400}{1050}{\nano\metre}, averaged over ten positions per sample, with shaded 1$\sigma$ bands.}
  \label{fig:methods-setup}
\end{figure}

Surface roughness was quantified by phase-shifting interferometry on a Sensofar Pl\textmu\ 2300 optical profiler with a 50$\times$/0.42 objective. All alloys exhibited root-mean-square (RMS) surface roughness of the same order of magnitude, with $S_q = \SI{12.1(1.6)}{\nano\metre}$ for FeNi, \SI{11.8(6)}{\nano\metre} for CoNi, \SI{12.7(1.3)}{\nano\metre} for CrFeNi, and \SI{14.2(6)}{\nano\metre} for CrCoNi, while CrMnFeCoNi was smoother, with a mean $S_q=\SI{4.1(1)}{\nano\metre}$.

\begin{table*}[!ht]
  \centering
  \caption{Average alloy compositions in at.\%, measured by energy-dispersive X-ray spectroscopy (EDX, values marked by *) and micro-X-ray fluorescence (\textmu-XRF).}
  \label{tab:edx-xrf}
  \footnotesize
  \setlength{\tabcolsep}{5pt}
  \renewcommand{\arraystretch}{1.15}
  \begin{tabular}{@{}lccccc@{}}
    \toprule
    {Alloy} & {Cr (at.\%)} & {Mn (at.\%)} & {Fe (at.\%)} & {Co (at.\%)} & {Ni (at.\%)} \\
    \midrule
    FeNi       & \textemdash & \textemdash & 50.8(4)* / 51.1(7) & \textemdash & 49.2(4)* / 48.9(7) \\
    CoNi       & \textemdash & \textemdash & \textemdash & 51.0(4)* / 48.9(2) & 49.0(4)* / 51.1(2) \\
    CrFeNi     & 34.9(1.4)* / 35.9(5) & \textemdash & 32.7(1.1)* / 32.6(2) & \textemdash & 32.4(2)* / 31.5(3) \\
    CrCoNi     & 34.7(1)* / 36.2(1.3) & \textemdash & \textemdash & 31.9(1.5)* / 30.4(1.0) & 33.4(1.4)* / 33.4(3) \\
    CrMnFeCoNi & 20.4(4)* & 19.9(6)* & 20.2(6)* & 20.2(7)* & 19.3(8)* \\
    \bottomrule
  \end{tabular}
\end{table*}

Chemical compositions were assessed by energy-dispersive X-ray spectroscopy (EDX) and micro-X-ray fluorescence (\textmu-XRF). EDX was carried out in a Tescan Lyra 3 scanning electron microscope (SEM) equipped with an Oxford Instruments detector, and \textmu-XRF with an EDAX Orbis PC SDD system (Rh anode X-ray tube, silicon drift detector) at \SI{45}{\kilo\volt} and \SI{400}{\micro\ampere} (two iterations, \SI{25}{\second} live time). For each method, atomic concentrations were averaged over multiple random positions per sample and are reported separately in Table~\ref{tab:edx-xrf}. All subsequent experiments and analyses were performed on these mirror-polished surfaces.

The average grain size was measured by the line-intercept method on backscattered electron (BSE) micrographs (Tescan Lyra 3 SEM, as above), yielding $d_{\mathrm{gr}} = \SI{287}{\micro\metre}$ for FeNi, \SI{646}{\micro\metre} for CoNi, \SI{484}{\micro\metre} for CrFeNi, \SI{596}{\micro\metre} for CrCoNi, and \SI{144}{\micro\metre} for CrMnFeCoNi, with no monotonic trend across the selected Cantor--Wu alloy series.

The Vickers microhardness was measured with a Falcon 500 hardness tester (Innovatest) at a load of \SI{0.5}{\kilo\gram} (HV0.5), averaged over five indents per alloy. The mean values were \SI{109(4)}{\HV} for FeNi, \SI{88(3)}{\HV} for CoNi, \SI{125(3)}{\HV} for CrFeNi, \SI{144(2)}{\HV} for CrCoNi, and \SI{150(6)}{\HV} for CrMnFeCoNi (Fig.~\ref{fig:methods-setup}a).
The grain-size contribution to the hardness was estimated from the Hall--Petch relation in hardness form,
\begin{equation}
  H_\mathrm{V} = H_{\mathrm{V}0} + k_\mathrm{H}\, d_{\mathrm{gr}}^{-1/2},
  \label{eq:hall-petch-hardness}
\end{equation}
using, as an approximation, literature Hall--Petch coefficients of related alloys, converted from their yield-stress form $k_\mathrm{y}$ to hardness units via $k_\mathrm{H}\approx 3k_\mathrm{y}/9.81$ (assuming $H_\mathrm{V}\approx 3\sigma_\mathrm{y}$, with $\sigma_\mathrm{y}$ the yield strength), namely $k_\mathrm{H} = \SI{50.2}{\HV\micro\metre^{1/2}}$ for austenitic steel~\cite{hirotaYieldStressDuplex2010} (applied to FeNi and CrFeNi), \SI{55.4}{\HV\micro\metre^{1/2}} for Ni-40Co (applied to CoNi) and \SI{81}{\HV\micro\metre^{1/2}} for CrCoNi~\cite{yoshidaFrictionStressHallPetch2017}, and \SI{151}{\HV\micro\metre^{1/2}} for CrMnFeCoNi~\cite{ottoInfluencesTemperatureMicrostructure2013}. The corresponding increments $\Delta H_\mathrm{V} = k_\mathrm{H}/\sqrt{d_{\mathrm{gr}}}$ are \SI{3.0}{\HV} for FeNi, \SI{2.2}{\HV} for CoNi, \SI{2.3}{\HV} for CrFeNi, \SI{3.3}{\HV} for CrCoNi, and \SI{12.6}{\HV} for CrMnFeCoNi.

In addition, optical reflectance spectra of the polished surfaces were recorded with a Filmetrics F20 spectral reflectometer (15$\times$ objective, spot diameter \SI{30}{\micro\metre}) at normal incidence from \SIrange{400}{1050}{\nano\metre}. Spectra were averaged over ten randomly selected positions per sample and calibrated against a silicon reference (Fig.~\ref{fig:methods-setup}b). For the ns wavelength of \SI{1064}{\nano\metre}, the reflectance was extrapolated from the flat spectral dispersion above \SI{1000}{\nano\metre}.

\subsection{Laser ablation experiments}
Single-pulse ablation thresholds were measured with separate femtosecond (fs) and nanosecond (ns) laser sources.

The fs experiments used a Light Conversion Pharos at $\lambda = \SI{1030}{\nano\metre}$, with a spectral width of \SI{7}{\nano\metre} and a beam quality of $M^2=\num{1.25}$. The pulse duration was set with an external pulse stretcher in Martinez configuration~\cite{martinezNegativeGroupvelocityDispersion1984} and measured by intensity autocorrelation (PulseCheck, APE) as \SI{400}{\femto\second}, assuming a Gaussian temporal pulse shape. The system was operated at \SI{500}{\hertz} with pulse-to-pulse energy fluctuations below \SI{1}{\percent} (standard deviation), and the incident fluence on the samples was adjusted using a combination of a half-wave plate and a polarizing beam splitter. Spot positioning was controlled by an $x$--$y$ translation stage, and the beam was focused onto the surface by a plano-convex lens ($f = \SI{100}{\milli\metre}$), yielding a beam waist radius (1/$\mathrm{e}^2$ intensity definition) of $w_0 = \SI{14.1(1)}{\micro\metre}$, measured with a focal beam profiler (Primes GmbH, MicroSpot Monitor) at an angle of incidence of $\theta = \ang{36.9}$. The beam was p-polarized with respect to the plane of incidence.

The ns experiments used a Coherent EasyMark 5 fibre laser at $\lambda = \SI{1064}{\nano\metre}$ with a beam quality of $M^2=\num{1.5}$. The pulse duration was \SI{102}{\nano\second} (FWHM), measured by detecting the laser radiation scattered from a white ceramic plate with a fast photodiode (Thorlabs DET10/A). Spot positioning was controlled by a galvanometer scanner combined with an f-theta lens ($f = \SI{160}{\milli\metre}$). The system was operated at \SI{100}{\kilo\hertz} with a scan speed of \SI{10}{\metre\per\second}, resulting in an inter-pulse distance of \SI{100}{\micro\metre} and a line spacing of \SI{100}{\micro\metre} between individual line scans. The pulse energy was adjusted via the diode pump current, and the average power was measured with a power meter placed \SI{80}{\milli\metre} above the focal plane. The beam waist radius on the sample was determined by the $D^2_\mathrm{abl}$ method~\cite{liuSimpleTechniqueMeasurements1982} as $w_0 = \SI{13(1)}{\micro\metre}$, and the focal position was identified by ablating at $z$-steps of \SI{250}{\micro\metre} and selecting the line with the most pronounced and circular craters. The beam was circularly polarized.

For the fs experiments, oblique incidence produced elliptical craters with a projected long-axis radius $w_x = w_0/\cos(\theta)$. The ablation threshold fluence and the effective beam waist were extracted by an adapted elliptical-area $D^2_\mathrm{abl}$ fit,
\begin{equation}
  A_\mathrm{abl} = \frac{\pi w_0^2}{2\cos\theta}\,\ln\!\left(\frac{F_0}{F_{\mathrm{thr}}}\right),
  \label{eq:elliptical-area-d2}
\end{equation}
where $A_\mathrm{abl}$ is the crater area measured by optical microscopy (Leitz Ergoplan, 50$\times$/0.85) from the long and short crater diameters averaged over ten craters per fluence level, $F_0$ is the peak fluence of the Gaussian beam, and $F_{\mathrm{thr}}$ is the ablation threshold fluence. For the ns experiments at normal incidence, crater diameters, averaged over ten craters per fluence level, were measured and analyzed with the standard $D^2_\mathrm{abl}$ method~\cite{liuSimpleTechniqueMeasurements1982},
\begin{equation}
  D^2_\mathrm{abl} = 2 w_0^2\,\ln\!\left(\frac{F_0}{F_{\mathrm{thr}}}\right),
  \label{eq:d2}
\end{equation}
where $D_\mathrm{abl}$ is the ablation crater diameter.

\subsection{First-principles calculations}
Density functional theory (DFT) calculations for all investigated alloys were performed using the fully relativistic, spin-polarized multiple-scattering Korringa--Kohn--Rostoker (KKR) Green's-function formalism, as implemented in the Munich spin-polarized relativistic KKR (SPR-KKR) program package~\cite{ebertCalculatingCondensedMatter2011}. Chemical disorder was treated in the coherent potential approximation (CPA)~\cite{faulknerCalculatingPropertiesCoherentpotential1980}, assuming ideal random solid solutions and neglecting atomic short-range order~\cite{woodgateCompositionalPhaseStability2022, woodgateInterplayMagnetismShortrange2023,chenDirectObservationChemical2021}. The exchange-correlation potential was treated in the local-density approximation (Vosko--Wilk--Nusair parametrization)~\cite{voskoAccurateSpindependentElectron1980}. For each composition, the face-centred cubic (fcc) lattice constant $a_\mathrm{fcc}$ was fixed to the experimentally reported room-temperature value (Table~\ref{tab:dft-inputs}). Self-consistent calculations used a dense $\mathbf{k}$-point mesh of 62$\times$62$\times$62, with the energy integration performed on a semicircular contour in the complex plane using 36 points and the angular-momentum expansion truncated at $\ell_\mathrm{max}=3$. Convergence tests for the transport calculations confirmed that $\ell_\mathrm{max}=3$ is sufficient.

The magnetic phase of each alloy was chosen according to its room-temperature character. The binary alloys FeNi and CoNi were treated as ferromagnets, whereas the ternary alloys CrFeNi and CrCoNi and the quinary alloy CrMnFeCoNi were treated as paramagnets through the disordered-local-moment (DLM) approach within CPA~\cite{pindorDisorderedLocalMoment1983}. For the ferromagnetic binary alloys, the temperature-dependent magnetization $m(T_\mathrm{ph})$, where $T_\mathrm{ph}$ denotes the phonon (lattice) temperature, was obtained from first principles in two steps. Interatomic exchange couplings $J_{ij}$ were extracted from the converged ferromagnetic SPR-KKR potential using the Liechtenstein formula~\cite{liechtensteinLocalSpinDensity1987}, resolved by species pair and averaged over coordination shells and concentration. These couplings were mapped onto a classical Heisenberg Hamiltonian, $H=-\sum_{i<j} J_{ij}\,\mathbf{e}_i\cdot\mathbf{e}_j$, on the fcc lattice (eight coordination shells) and solved by atomistic spin Monte-Carlo simulations with the VAMPIRE package~\cite{evansAtomisticSpinModel2014} on a \SI{10}{\nano\metre} cube ($\approx\num{88000}$ atoms). The Curie temperature was obtained from a critical power-law fit $m(T_\mathrm{ph})=A\,(1-T_\mathrm{ph}/T_\mathrm{C})^{\beta}$ with $\beta=0.365$, giving $T_\mathrm{C}=\SI{735}{\kelvin}$ for FeNi and \SI{909}{\kelvin} for CoNi. The FeNi value agrees with the experimental Curie temperature to within approximately \SI{6}{\percent} (\SI{780}{\kelvin}~\cite{jinThermophysicalPropertiesNicontaining2017}, \SI{762}{\kelvin}~\cite{sharmaCoexistingMagneticInteraction2024}), while for CoNi the calculation underestimates the measured value of \SI{1117}{\kelvin}~\cite{jinThermophysicalPropertiesNicontaining2017} by approximately \SI{19}{\percent}.

\begin{table*}[t]
  \centering
  \caption{Parameters used for the SPR-KKR calculations. Lattice constants are taken from Ref.~\cite{muUncoveringElectronScattering2019} and marked with * when taken from Ref.~\cite{laplancheElasticModuliThermal2018}. The alloy mass-enhancement parameter, $\lambda$, is the concentration-weighted average of the elemental values (Cr~0.131, Mn~0.364, Fe~0.270, Co~0.197, Ni~0.084) from Ref.~\cite{papaconstantopoulosCalculationsSuperconductingProperties1977a}. Alloy Debye temperatures, $\Theta_\mathrm{D}$, are elastic estimates obtained from the shear modulus $G$ and bulk modulus $K$, taken from Ref.~\cite{singhElasticConstantsEquiatomic2023} (FeNi), Ref.~\cite{laplancheElasticModuliThermal2018} (CoNi, CrFeNi, CrCoNi), and Ref.~\cite{laplancheTemperatureDependenciesElastic2015} (CrMnFeCoNi). The Eliashberg moment is $\lambda\langle\omega^2\rangle=\tfrac{1}{2}\lambda(k_\mathrm{B}\Theta_\mathrm{D})^2$, and $N$ is the number of valence electrons per atom.}
  \label{tab:dft-inputs}
  \footnotesize
  \setlength{\tabcolsep}{5pt}
  \renewcommand{\arraystretch}{1.15}
  \sisetup{table-number-alignment=center, table-text-alignment=center, detect-weight=true, detect-family=true}
  \begin{tabular}{@{}l
      S[table-format=1.4]
      S[table-format=1.3]
      S[table-format=3.0]
      S[table-format=3.0]
      c
      S[table-format=1.3]
      S[table-format=3.1]
    S[table-format=1.3]@{}}
    \toprule
    {Alloy} & {$a_{\mathrm{fcc}}$ (\si{\Angstrom})} & {$\rho_\mathrm{m}$ (\si{\gram\per\centi\metre\cubed})} & {$G$ (\si{\giga\pascal})} & {$K$ (\si{\giga\pascal})} & {$\Theta_\mathrm{D}$ (\si{\kelvin})} & {$\lambda$} & {$\lambda\langle\omega^2\rangle$ (\si{\milli\electronvolt\squared})} & {$N$} \\
    \midrule
    FeNi       & 3.5826 & 8.272 & 74 & 99  & 434 & 0.177 & 123.9 & 9.000 \\
    CoNi       & 3.5345 & 8.847 & 76 & 534 & 446 & 0.141 & 103.6 & 9.500 \\
    CrFeNi     & 3.590* & 7.969 & 79 & 104 & 457 & 0.162 & 125.6 & 7.996 \\
    CrCoNi     & 3.5590 & 8.331 & 90 & 208 & 488 & 0.137 & 121.5 & 8.330 \\
    CrMnFeCoNi & 3.5991 & 7.990 & 81 & 137 & 464 & 0.209 & 167.5 & 8.000 \\
    \bottomrule
  \end{tabular}
\end{table*}

The electronic density of states $n(\varepsilon)$ was evaluated in an energy window of $\pm\SI{15}{\electronvolt}$ around the Fermi level $\varepsilon_\mathrm{F}$. This window safely accommodates the smearing of the Fermi--Dirac distribution $f(\varepsilon,\mu,T_\mathrm{e})$ ($\approx 5k_\mathrm{B}T_\mathrm{e}$) for electron temperatures up to \SI{30}{\kilo\kelvin}, while the present analysis was restricted to $T_\mathrm{e}\le\SI{20}{\kilo\kelvin}$.

The chemical potential $\mu(T_\mathrm{e})$ was determined numerically from charge conservation,
\begin{equation}
  N = \int_{-\infty}^{\infty} n(\varepsilon)\, f(\varepsilon,\mu,T_\mathrm{e})\,\mathrm{d}\varepsilon,
  \label{eq:charge-neutrality}
\end{equation}
with $N$ the total number of valence electrons per atom.

The electronic heat capacity $C_\mathrm{e}(T_\mathrm{e})$ was computed via
\begin{equation}
  \begin{aligned}
    C_\mathrm{e}(T_\mathrm{e}) &= \frac{1}{k_\mathrm{B} T_\mathrm{e}^2}\left(M_2-\frac{M_1^2}{M_0}\right),\\
    M_n &= \int_{-\infty}^{\infty} n(\varepsilon)\,(\varepsilon-\mu)^n\,f\,(1-f)\,\mathrm{d}\varepsilon,
  \end{aligned}
  \label{eq:ce_te}
\end{equation}
where $f=f(\varepsilon,\mu,T_\mathrm{e})$. A linear fit $C_\mathrm{e}(T_\mathrm{e})=\gamma\,T_\mathrm{e}$ up to $T_\mathrm{e}=\SI{2000}{\kelvin}$ yielded the Sommerfeld parameter, $\gamma$.

The electron--phonon coupling factor $G_\mathrm{ep}(T_\mathrm{e})$ was computed via the Allen formalism~\cite{allenTheoryThermalRelaxation1987,linElectronphononCouplingElectron2008},
\begin{equation}
  G_\mathrm{ep}(T_\mathrm{e})=\frac{\pi\hbar k_\mathrm{B}\,\lambda\langle\omega^2\rangle}{n(\varepsilon_\mathrm{F})}\int_{-\infty}^{\infty} n^2(\varepsilon)\left(-\frac{\partial f}{\partial\varepsilon}\right)\,\mathrm{d}\varepsilon,
  \label{eq:g_te}
\end{equation}
where $\lambda\langle\omega^2\rangle$ denotes the second moment of the Eliashberg function~\cite{allenTheoryThermalRelaxation1987}. The second phonon moment $\langle\omega^2\rangle$ was approximated from the Debye temperature $\Theta_\mathrm{D}$ (assuming a Debye solid)~\cite{papaconstantopoulosCalculationsSuperconductingProperties1977a} as $\langle\omega^2\rangle = (k_\mathrm{B}\Theta_\mathrm{D})^2/2$, with the resulting $\lambda\langle\omega^2\rangle$ given in Table~\ref{tab:dft-inputs}. For multicomponent alloys, the mass-enhancement parameter was approximated by a concentration-weighted average, $\lambda = \sum_i x_i\,\lambda_i$, with elemental mass-enhancement parameters $\lambda_i$ taken from Ref.~\cite{papaconstantopoulosCalculationsSuperconductingProperties1977a} (Table~\ref{tab:dft-inputs}) and $x_i$ the atomic fraction of element $i$~\cite{winterTemperaturedependentDeterminationElectron2016}. The Debye temperature was not taken as a concentration-weighted elemental average but from the measured polycrystalline elastic moduli. The shear modulus $G$, bulk modulus $K$, and mass density $\rho_\mathrm{m}$ define the transverse, longitudinal, and mean sound velocities~\cite{moruzziCalculatedThermalProperties1988},
\begin{equation}
  \begin{gathered}
    v_\mathrm{t} = \sqrt{G/\rho_\mathrm{m}},\qquad v_\mathrm{l} = \sqrt{(K + \tfrac{4}{3}G)/\rho_\mathrm{m}},\\
    v_\mathrm{m} = \left[\frac{1}{3}\left(\frac{2}{v_\mathrm{t}^{3}} + \frac{1}{v_\mathrm{l}^{3}}\right)\right]^{-1/3},
  \end{gathered}
  \label{eq:sound-velocities}
\end{equation}
from which the Debye temperature follows as
\begin{equation}
  \Theta_\mathrm{D} = \frac{\hbar}{k_\mathrm{B}}\left(\frac{24\pi^2}{a_\mathrm{fcc}^3}\right)^{1/3} v_\mathrm{m}.
  \label{eq:debye-elastic}
\end{equation}
The shear and bulk moduli used, with the corresponding references, are reported in Table~\ref{tab:dft-inputs}.

The longitudinal DC electrical conductivity was obtained in the Kubo--Greenwood approach~\cite{kuboStatisticalMechanicalTheoryIrreversible1957,butlerTheoryElectronicTransport1985,banhartOpticalConductivityDisordered1999} according to
\begin{equation}
  \sigma_{\alpha\alpha}(\varepsilon_\mathrm{F}) = \frac{\hbar}{V\pi}\,\operatorname{Tr}\,\Big\langle \hat{j}_{\alpha}\,\operatorname{Im} \mathcal{G}^+(\varepsilon_\mathrm{F})\,\hat{j}_{\alpha}\,\operatorname{Im} \mathcal{G}^+(\varepsilon_\mathrm{F}) \Big\rangle_\mathrm{c},
  \label{eq:kubo_greenwood}
\end{equation}
where $V$ is the unit cell volume, $\hat{j}_{\alpha}$ is the current-density operator, $\mathcal{G}^+(\varepsilon_\mathrm{F})$ is the retarded Green's function at the Fermi level, and $\langle\cdots\rangle_\mathrm{c}$ denotes the configurational average within CPA, including vertex corrections~\cite{butlerTheoryElectronicTransport1985,kodderitzschLinearResponseKuboBastin2015}. For the cubic alloys considered here, the diagonal components are equal, and we use $\sigma\equiv\sigma_{xx}=\sigma_{yy}=\sigma_{zz}$. The electronic thermal conductivity $\kappa_\mathrm{e}$ then follows from the Wiedemann--Franz law, $\kappa_\mathrm{e} = L_0\,T_\mathrm{ph}/\rho$, with $\rho = 1/\sigma$ and the Lorenz number $L_0$.

For the Kubo--Greenwood conductivity calculations, convergence with respect to Brillouin-zone sampling was reached for a 54$\times$54$\times$54 $\mathbf{k}$-point mesh. Finite lattice temperatures were treated within the alloy-analogy model~\cite{ebertCalculatingLinearresponseFunctions2015} by introducing temperature-dependent static atomic displacements (sampled equally weighted over 14 uniformly distributed directions) with a root-mean-square amplitude $\sqrt{\langle u^2\rangle}$ obtained from the Debye model,
\begin{equation}
  \langle u^2\rangle = \frac{9\hbar^2}{M_\mathrm{at} k_\mathrm{B} \Theta_\mathrm{D}}\left[\frac{1}{4} + \left(\frac{T_\mathrm{ph}}{\Theta_\mathrm{D}}\right)^2 \int_{0}^{\Theta_\mathrm{D}/T_\mathrm{ph}} \frac{t}{\mathrm{e}^t-1} \mathrm{d}t\right],
  \label{eq:debye-msd}
\end{equation}
where $M_\mathrm{at}$ is the composition-averaged atomic mass. Thermal spin disorder was included on the same footing within the alloy-analogy model~\cite{ebertCalculatingLinearresponseFunctions2015}, with the local-moment orientations sampled over 18 polar ($\theta$) and 8 azimuthal ($\phi$) directions. For the ferromagnetic binary alloys, the orientational distribution was set by the calculated magnetization $m(T_\mathrm{ph})$, so that the spin disorder grows with temperature and the magnetization vanishes at $T_\mathrm{ph}=T_\mathrm{C}$. For the paramagnetic alloys, the average over magnetic disorder was treated using the DLM picture at all temperatures~\cite{redkaInterplayDisorderElectronic2024}.

The optical conductivity was computed from the frequency-dependent Kubo--Greenwood expression~\cite{banhartOpticalConductivityDisordered1999,perlovInitioCalculationOptical2004}, yielding the absorptive part
\begin{equation}
  \sigma^{(1)}_{\alpha\alpha}(\omega) = \frac{1}{\pi V\omega}\!\int_{\varepsilon_\mathrm{F}-\hbar\omega}^{\varepsilon_\mathrm{F}}\!\!
  \operatorname{Tr}\big\langle \hat{j}_{\alpha}\,\operatorname{Im}\mathcal{G}^+(\varepsilon)\,\hat{j}_{\alpha}\operatorname{Im}\mathcal{G}^+(\varepsilon{+}\hbar\omega)\big\rangle_\mathrm{c}\,\mathrm{d}\varepsilon.
  \label{eq:optical_sigma1}
\end{equation}
The dispersive part $\sigma^{(2)}(\omega)$ was obtained from a Kramers--Kronig transformation, giving the complex conductivity $\sigma(\omega)=\sigma^{(1)}(\omega)+\mathrm{i}\sigma^{(2)}(\omega)$. $\mathbf{k}$-point convergence (19$\times$19$\times$19) was verified by confirming that the $\omega\to 0$ limit of $\sigma(\omega)$ matches the independently computed static conductivity (as described above). The complex dielectric function was then computed as
\begin{equation}
  \varepsilon(\omega) = 1 + \frac{\mathrm{i}\,\sigma(\omega)}{\varepsilon_0\,\omega},
  \label{eq:epsilon_from_sigma}
\end{equation}
and the reflectance and absorbance were obtained from $\varepsilon(\omega)$ via the Fresnel equations.

\section{Results}
\subsection{Ablation thresholds and morphology}

\begin{figure*}[!ht]
  \centering
  \includegraphics[width=\textwidth]{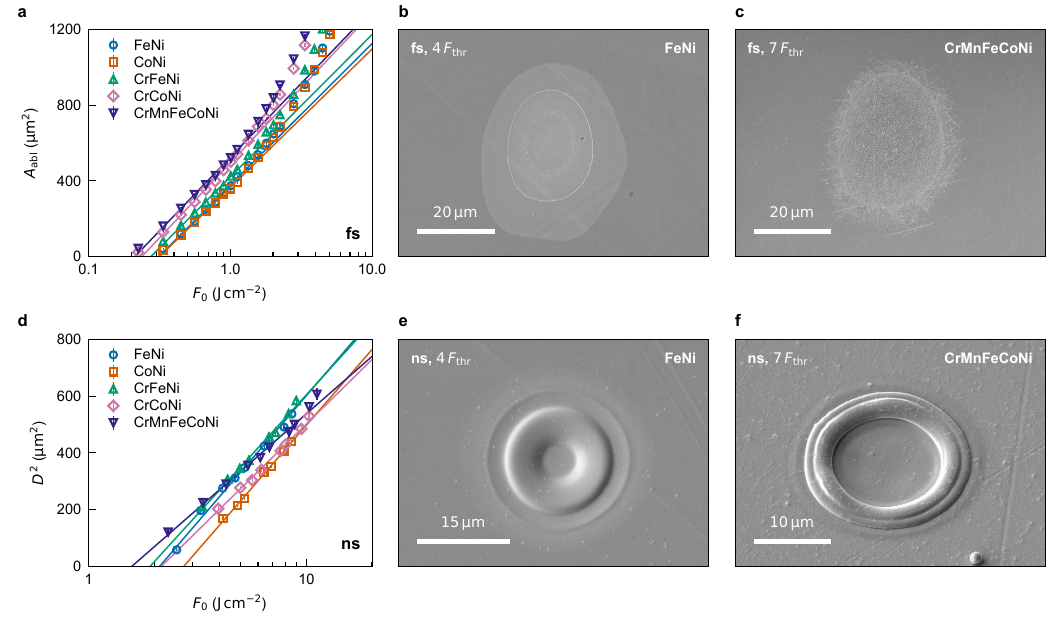}
  \caption{Single-pulse ablation thresholds and crater morphologies of the selected Cantor--Wu alloys (FeNi, CoNi, CrFeNi, CrCoNi, CrMnFeCoNi), grouped by pulse duration: femtosecond (fs, top row) and nanosecond (ns, bottom row). \textbf{a,} Femtosecond crater area, $A_\mathrm{abl}$ ($\lambda=\SI{1030}{\nano\metre}$, $\tau_\mathrm{p}=\SI{400}{\femto\second}$, angle of incidence $\theta=\ang{36.9}$), versus incident peak fluence, $F_0$, for all five alloys (symbols, measured data; lines, elliptical-area $D^2_\mathrm{abl}$ fits that yield the threshold fluence, $F_{\mathrm{thr}}$; colours in the legend). \textbf{b,} FeNi and \textbf{c,} CrMnFeCoNi craters for the fs pulse duration, imaged at $4\,F_{\mathrm{thr}}$ and $7\,F_{\mathrm{thr}}$. \textbf{d,} Nanosecond squared crater diameter, $D^2$ ($\lambda=\SI{1064}{\nano\metre}$, $\tau_\mathrm{p}=\SI{102}{\nano\second}$, normal incidence), versus $F_0$ (symbols, data; lines, standard $D^2_\mathrm{abl}$ fits). \textbf{e,} FeNi and \textbf{f,} CrMnFeCoNi craters for the ns pulse duration at the same fluence multiples. Panels \textbf{b}, \textbf{c}, \textbf{e}, \textbf{f} are greyscale scanning electron microscopy (SEM) micrographs with scale bars of \SI{20}{\micro\metre} (b, c), \SI{15}{\micro\metre} (e), and \SI{10}{\micro\metre} (f).}
  \label{fig:thresholds}
\end{figure*}

The resulting elliptical craters of the fs experiments (Methods) were analyzed with an adapted elliptical-area $D^2_\mathrm{abl}$ fit. Figure~\ref{fig:thresholds}a shows the crater areas, $A_\mathrm{abl}$, versus incident peak fluence, $F_0$, with the corresponding fits. The fitted beam waist varies by less than \SI{0.5}{\micro\metre} across compositions, and its mean of \SI{13.0(2)}{\micro\metre} matches the short-axis waist of \SI{14.1(1)}{\micro\metre} measured independently with the focal beam profiler (Methods). The incident threshold fluences (Table~\ref{tab:thresholds}) are nearly identical for FeNi and CoNi. Adding Cr lowers the threshold, and CrMnFeCoNi reaches the lowest value. The incident threshold thus decreases monotonically with chemical complexity, by \SI{32}{\percent} from FeNi to CrMnFeCoNi. These values are consistent with the reported range for FeNi~\cite{chenTimeresolvedProbingModeling2025} and CrMnFeCoNi~\cite{redkaSubpicosecondSinglepulseLaser2021} and with literature on 3$d$ transition metals and alloys~\cite{mannionEffectDamageAccumulation2004,semaltianosFemtosecondLaserAblation2009,thomaeDecipheringDrivingMechanisms2026} across moderate variations in pulse duration.

\begin{table*}[t]
  \centering
  \caption{Measured single-pulse ablation threshold fluences of the Cantor--Wu alloys for femtosecond ($\lambda = \SI{1030}{\nano\metre}$, $\tau_\mathrm{p} = \SI{400}{\femto\second}$, $\theta = \ang{36.9}$, p-pol.) and nanosecond ($\lambda = \SI{1064}{\nano\metre}$, $\tau_\mathrm{p} = \SI{102}{\nano\second}$, normal incidence) pulse durations. Incident thresholds $F_{\mathrm{thr}}$ are obtained from the $D^2$-fits; absorbed thresholds follow as $F_{\mathrm{abs}} = F_{\mathrm{thr}}(1 - R)$, with the reflectance $R$ measured at normal incidence at \SI{1030}{\nano\metre}. Owing to the flat spectral dispersion of the alloys, the same absorptance is used at the ns wavelength of \SI{1064}{\nano\metre}. Uncertainties are 1$\sigma$ fit uncertainties, propagated to $F_{\mathrm{abs}}$.}
  \label{tab:thresholds}
  \footnotesize
  \setlength{\tabcolsep}{6pt}
  \renewcommand{\arraystretch}{1.15}
  \begin{tabular}{@{}lcccc@{}}
    \toprule
    {Alloy} & {$F_{\mathrm{thr}}$ fs (\si{\joule\per\centi\metre\squared})} & {$F_{\mathrm{abs}}$ fs (\si{\joule\per\centi\metre\squared})} & {$F_{\mathrm{thr}}$ ns (\si{\joule\per\centi\metre\squared})} & {$F_{\mathrm{abs}}$ ns (\si{\joule\per\centi\metre\squared})}\\
    \midrule
    FeNi & \SI{0.312(4)}{\joule\per\centi\metre\squared} & \SI{0.095(1)}{\joule\per\centi\metre\squared} & \SI{2.12(2)}{\joule\per\centi\metre\squared} & \SI{0.649(6)}{\joule\per\centi\metre\squared}\\
    CoNi & \SI{0.317(5)}{\joule\per\centi\metre\squared} & \SI{0.087(1)}{\joule\per\centi\metre\squared} & \SI{2.74(2)}{\joule\per\centi\metre\squared} & \SI{0.748(5)}{\joule\per\centi\metre\squared}\\
    CrFeNi & \SI{0.275(7)}{\joule\per\centi\metre\squared} & \SI{0.076(2)}{\joule\per\centi\metre\squared} & \SI{1.92(2)}{\joule\per\centi\metre\squared} & \SI{0.528(6)}{\joule\per\centi\metre\squared}\\
    CrCoNi & \SI{0.237(4)}{\joule\per\centi\metre\squared} & \SI{0.061(1)}{\joule\per\centi\metre\squared} & \SI{2.16(2)}{\joule\per\centi\metre\squared} & \SI{0.559(5)}{\joule\per\centi\metre\squared}\\
    CrMnFeCoNi & \SI{0.212(4)}{\joule\per\centi\metre\squared} & \SI{0.064(1)}{\joule\per\centi\metre\squared} & \SI{1.58(3)}{\joule\per\centi\metre\squared} & \SI{0.476(9)}{\joule\per\centi\metre\squared}\\
    \bottomrule
  \end{tabular}
\end{table*}

The crater morphologies were imaged at fixed fluence multiples, $4\,F_{\mathrm{thr}}$ for FeNi and $7\,F_{\mathrm{thr}}$ for CrMnFeCoNi. For the fs pulse duration, the FeNi crater (Fig.~\ref{fig:thresholds}b) shows a pronounced crater edge, characteristic of spallation~\cite{inogamovNanospallationInducedUltrashort2008,zhigileiAtomisticModelingShort2009,linUltrafastMeltingSpallation2025}. Close to the crater center, a concentric circular region characteristic of phase explosion appears~\cite{ohFemtosecondLaserAblation2007,winterUltrashortSinglepulseLaser2021,redkaSubpicosecondSinglepulseLaser2021}. In addition, the spallation rim is surrounded by an annular zone of resolidified melt, corresponding to the fluence range between surface melting and the spallation onset, as identified previously for iron--nickel targets~\cite{chenTimeresolvedProbingModeling2025} and for CrMnFeCoNi and stainless steel AISI 304~\cite{redkaSubpicosecondSinglepulseLaser2021}. CrMnFeCoNi (Fig.~\ref{fig:thresholds}c) shows a comparable combination of spallation and phase explosion, but additionally exhibits intensified nano-scratches, both inside and outside the ablation crater. No such molten rim is resolved for CrMnFeCoNi. These nano-scratches are a leftover from the polishing routine and are not captured by the RMS surface roughness. Field enhancement at such nanostructures~\cite{rudenkoLightAbsorptionSurface2019,thomaeSurfaceRoughnessCrater2024} results in a locally higher fluence interacting with the material, so that the scratches are deepened even outside the crater, where the incident fluence stays below the threshold. The same mechanism has been reported for sub-picosecond ablation of CrMnFeCoNi~\cite{redkaSubpicosecondSinglepulseLaser2021} and is reproduced by finite-difference time-domain (FDTD) simulations of the near field at such surface defects~\cite{rudenkoLightAbsorptionSurface2019} and by multi-scale FDTD--TTM modelling linking surface roughness to localized fluence enhancement and hot-spot formation~\cite{thomaeSurfaceRoughnessCrater2024}.

For the ns pulse duration, the same procedure was applied at $\lambda=\SI{1064}{\nano\metre}$ and $\tau_\mathrm{p}=\SI{102}{\nano\second}$ at normal incidence, with the threshold fluences determined from the measured crater diameters by the standard $D^2_\mathrm{abl}$ method (Fig.~\ref{fig:thresholds}d). The fitted beam waists average \SI{13.2(8)}{\micro\metre}, consistent with the independent reference value $w_0=\SI{13}{\micro\metre}$ (Methods), and their larger spread across compositions of \SI{1.9}{\micro\metre} reflects the finite accuracy of the focal-position determination for each sample. Here, CoNi has the highest incident threshold of \SI{2.74(2)}{\joule\per\centi\metre\squared} and CrMnFeCoNi the lowest (Table~\ref{tab:thresholds}), a \SI{42}{\percent} reduction between CoNi and CrMnFeCoNi that exceeds the \SI{32}{\percent} decrease measured for the fs pulse duration. In general, the measured values are of the same order as reported ns-ablation fluences for stainless steel~\cite{khalilStudyExperimentalNumerical2005,naghilouFemtoNanosecondPulse2017} and consistent with the established increase of the ablation threshold with pulse duration~\cite{mustafaInfluencePulseDuration2020}. Overall, the composition-dependent trend aligns with the fs regime, with the exception of CoNi, which exceeds FeNi by \SI{30}{\percent} under ns pulses rather than matching it.

For the ns pulse duration, neither alloy develops a sharp crater edge. The FeNi crater (Fig.~\ref{fig:thresholds}e) is circular and smooth with a central resolidified bulge, a soft doughnut-shaped profile caused by recoil-pressure-driven displacement of the molten surface~\cite{kornerPhysicalMaterialAspects1996,mustafaInfluencePulseDuration2020}. The CrMnFeCoNi crater (Fig.~\ref{fig:thresholds}f) develops a pronounced peripheral rim of resolidified melt, reflecting melt expulsion during the long pulse duration and the higher relative fluence applied to CrMnFeCoNi than to FeNi, while the crater center remains smooth and round. The scratch-enhanced ablation seen for the fs pulse duration is absent, because the extended melting smooths over the surface scratches, as also observed once the pulse duration is extended into the picosecond range~\cite{redkaControlUltrafastLaser2022}. The generally round, rim-free morphology contrasts with the spallation-formed edge of the fs craters, and the resulting advantage of ultrashort pulses---a clean and well-defined crater rim---would become most apparent under multi-pulse irradiation~\cite{mommaShortpulseLaserAblation1996}.

\subsection{Electronic structure and transport}
\label{sec:results-dft}

\begin{figure*}[!ht]
  \centering
  \includegraphics[width=\textwidth]{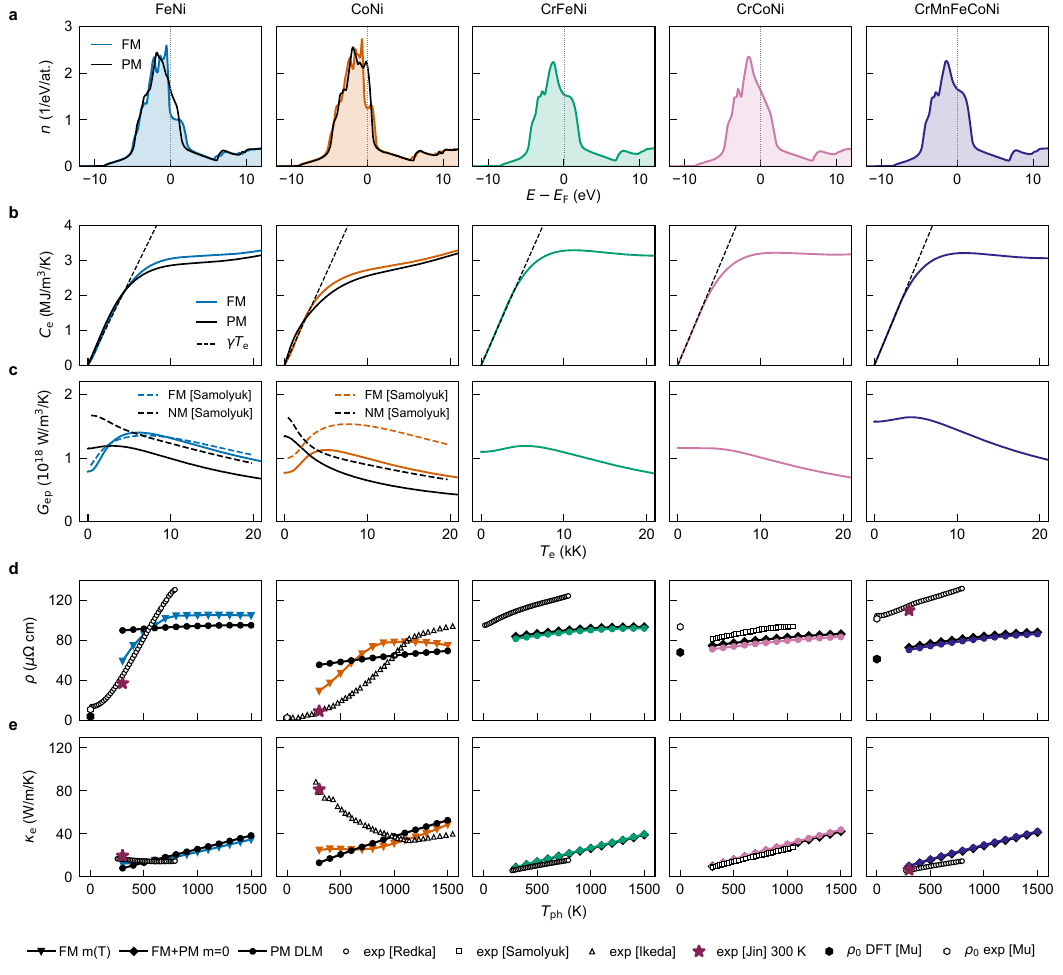}
  \caption{Electronic structure and transport properties from DFT of the Cantor--Wu alloys (columns: FeNi, CoNi, CrFeNi, CrCoNi, CrMnFeCoNi). Each alloy is shown in its room-temperature magnetic phase (ferromagnetic, FM, for FeNi and CoNi, and paramagnetic disordered local moment, PM/DLM, for CrFeNi, CrCoNi, and CrMnFeCoNi). For the two binary alloys, the PM/DLM result is overlaid in black. \textbf{a,} Total density of states, $n(\varepsilon)$, around the Fermi level, $\varepsilon_\mathrm{F}$. \textbf{b,} Electronic heat capacity, $C_\mathrm{e}(T_\mathrm{e})$, with the low-temperature Sommerfeld fit, $\gamma T_\mathrm{e}$ (dashed). \textbf{c,} Electron--phonon coupling factor $G_\mathrm{ep}(T_\mathrm{e})$ from the Allen formalism (Methods, Eq.~\eqref{eq:g_te}). For FeNi and CoNi, the first-principles reference calculations of Samolyuk \textit{et al.}~\cite{samolyukElectronPhononCoupling2016} (magnetic and non-magnetic) are shown as dashed lines. \textbf{d,} Electrical resistivity, $\rho(T_\mathrm{ph})$, from the linear-response calculations for the different magnetic configurations (FM with $m(T_\mathrm{ph})$, FM potential with $m=0$, PM/DLM), compared with experimental data~\cite{redkaInterplayDisorderElectronic2024,ikedaElectricalResistivityNickel1988,samolyukTemperatureDependentElectronic2018,jinTailoringPhysicalProperties2016} and with calculated and measured residual resistivities, $\rho_0$, from Ref.~\cite{muUncoveringElectronScattering2019}. \textbf{e,} Electronic thermal conductivity, $\kappa_\mathrm{e} = L_0 T_\mathrm{ph}/\rho$ (Wiedemann--Franz transform of \textbf{d}). Experimental points below \SI{250}{\kelvin} are omitted, where residual-resistivity-dominated $\rho$ renders the transform unphysical.}
  \label{fig:Figure_DFT}
\end{figure*}

Within the outlined first-principles electronic structure framework (Methods), the electronic heat capacity, $C_\mathrm{e}$, the electron--phonon coupling, $G_\mathrm{ep}$, and the electronic thermal conductivity, $\kappa_\mathrm{e}$ were determined for all investigated Cantor--Wu alloys. The resulting data are summarized in Fig.~\ref{fig:Figure_DFT}. FeNi and CoNi were treated in the ferromagnetic (FM) phase, the ternary alloys as well as CrMnFeCoNi in the paramagnetic (PM) DLM description~\cite{pindorDisorderedLocalMoment1983,gyorffyFirstprinciplesTheoryFerromagnetic1985}, in line with the finding that these alloys are paramagnetic at room temperature~\cite{jinTailoringPhysicalProperties2016,schneeweissMagneticPropertiesCrMnFeCoNi2017,billingtoBulkElementSpecificMagnetism2020,redkaInterplayDisorderElectronic2024}. For the binary alloys, results for the PM phase are additionally shown to isolate the influence of magnetic order.

Figure~\ref{fig:Figure_DFT}a shows the total density of states (DOS), $n(\varepsilon)$. All compositions exhibit a narrow $d$ band complex---associated with weakly dispersive $3d$ bands---typical of magnetic transition metals~\cite{stauntonElectronicStructureMagnetic1994}, spanning an energy range of approximately \SI{7}{\electronvolt} around $\varepsilon_\mathrm{F}$. The center of the occupied $d$ band relative to the Fermi level (its first spectral moment), $\varepsilon_d$, varies weakly across the series, amounting to \SI{-2.06}{\electronvolt} (FeNi), \SI{-2.17}{\electronvolt} (CoNi), \SI{-2.05}{\electronvolt} (CrFeNi), \SI{-2.05}{\electronvolt} (CrCoNi), and \SI{-1.99}{\electronvolt} (CrMnFeCoNi).
Despite the overall similarity of $n(\varepsilon)$, pronounced differences emerge at $\varepsilon_\mathrm{F}$. In ferromagnetic FeNi and CoNi, the exchange splitting between the spin-up and spin-down partial DOS pushes the majority $d$ states below $\varepsilon_\mathrm{F}$ and reduces the Fermi-level DOS to $n(\varepsilon_\mathrm{F}) = \SI{1.11}{\per\electronvolt}$ and \SI{1.24}{\per\electronvolt} per atom, respectively, compared with \SI{1.54}{\per\electronvolt} to \SI{1.66}{\per\electronvolt} per atom for the paramagnetic Cr-containing alloys, a reduction of approximately \SI{30}{\percent}. Relative to their own PM phases (\SI{1.62}{\per\electronvolt} and \SI{2.17}{\per\electronvolt} per atom), the reduction reaches \SI{43}{\percent} for CoNi. The Fermi-level DOS of the binary alloys resides predominantly in the minority-spin channel, and chemical and magnetic disorder scattering is largely confined to this channel, while the majority-spin states maintain a comparatively sharp quasiparticle Fermi surface~\cite{muUncoveringElectronScattering2019}. The corresponding local magnetic moments are \SI{2.52}{\muB} (Fe) and \SI{0.68}{\muB} (Ni) in FeNi, and \SI{1.65}{\muB} (Co) and \SI{0.66}{\muB} (Ni) in CoNi. In CrFeNi, CrCoNi, and CrMnFeCoNi, the Cr, Co and Ni moments collapse during the DFT self-consistency cycle, as has been extensively reported for DLM calculations within the CPA~\cite{pindorDisorderedLocalMoment1983,muHiddenMnMagneticmoment2019, woodgateCompositionalPhaseStability2022, woodgateInterplayMagnetismShortrange2023}. In CrFeNi the Fe moment is \SI{1.60}{\muB}, while in CrMnFeCoNi the moments are \SI{1.66}{\muB} (Fe) and \SI{1.29}{\muB} (Mn).

The electronic heat capacity as a function of the electronic temperature, $C_\mathrm{e}(T_\mathrm{e})$, is shown in Fig.~\ref{fig:Figure_DFT}b together with the low-temperature Sommerfeld fit $C_\mathrm{e}=\gamma T_\mathrm{e}$ (Methods, Eq.~\eqref{eq:ce_te}). The fitted Sommerfeld parameters increase from $\gamma = \SI{481}{\joule\per\metre\cubed\per\kelvin\squared}$ (FeNi) and \SI{520}{\joule\per\metre\cubed\per\kelvin\squared} (CoNi) to \SI{534}{\joule\per\metre\cubed\per\kelvin\squared} (CrFeNi), \SI{563}{\joule\per\metre\cubed\per\kelvin\squared} (CrCoNi), and \SI{568}{\joule\per\metre\cubed\per\kelvin\squared} (CrMnFeCoNi), following the trend in $n(\varepsilon_\mathrm{F})$. The linear regime is restricted to $T_\mathrm{e} \lesssim \SI{2}{\kilo\kelvin}$. At higher electronic temperatures the Fermi--Dirac smearing samples the strongly structured $d$ band and all alloys converge to a weakly increasing plateau of approximately \SI{3}{\mega\joule\per\metre\cubed\per\kelvin} above $T_\mathrm{e} \approx \SI{10}{\kilo\kelvin}$. The FM and PM results for the binary alloys differ only weakly.

The TTM energy-transfer rate from the laser-excited electron subsystem to the lattice is quantified by the electron--phonon coupling factor, $G_\mathrm{ep}(T_\mathrm{e})$, computed via the Allen formalism (Methods, Eq.~\eqref{eq:g_te})~\cite{allenTheoryThermalRelaxation1987,linElectronphononCouplingElectron2008}. In this approach, $G_\mathrm{ep}(T_\mathrm{e})$ is controlled by the DOS in the thermally broadened window around $\varepsilon_\mathrm{F}$, weighted by the derivative of the Fermi--Dirac function, and by the Eliashberg spectral moment $\lambda\langle\omega^2\rangle$ (Table~\ref{tab:dft-inputs}), obtained from concentration-weighted elemental mass-enhancement parameters and elastic Debye temperatures~\cite{papaconstantopoulosCalculationsSuperconductingProperties1977a,winterTemperaturedependentDeterminationElectron2016} (Methods). The resulting coupling factors, $G_\mathrm{ep}(T_\mathrm{e})$, up to \SI{20}{\kilo\kelvin} are shown in Fig.~\ref{fig:Figure_DFT}c. At low $T_\mathrm{e}$, the ferromagnetic binary alloys exhibit markedly smaller coupling, with $G_\mathrm{ep}$ of \SI{0.79e18}{\watt\per\metre\cubed\per\kelvin} (FeNi) and \SI{0.77e18}{\watt\per\metre\cubed\per\kelvin} (CoNi), consistent with their reduced DOS at $\varepsilon_\mathrm{F}$. The paramagnetic alloys yield larger values of \SI{1.10e18}{\watt\per\metre\cubed\per\kelvin} (CrFeNi), \SI{1.16e18}{\watt\per\metre\cubed\per\kelvin} (CrCoNi), and \SI{1.57e18}{\watt\per\metre\cubed\per\kelvin} (CrMnFeCoNi). The quinary alloy combines the highest $n(\varepsilon_\mathrm{F})$ with the largest Eliashberg moment ($\lambda\langle\omega^2\rangle = \SI{167.5}{\milli\electronvolt\squared}$, dominated by the large elemental $\lambda$ of Mn) and therefore shows the strongest coupling over the entire $T_\mathrm{e}$ range. The temperature dependence of $G_\mathrm{ep}$ reflects the DOS sampled by the thermally broadening Fermi--Dirac window in Eq.~\eqref{eq:g_te}. As this window broadens across the $d$-band structure, $G_\mathrm{ep}(T_\mathrm{e})$ increases in FeNi and CoNi up to $T_\mathrm{e} \approx \SI{5}{\kilo\kelvin}$, after which the coupling decreases monotonically for all alloys, reflecting the finite extent of the unoccupied $d$-band states of about \SI{2.5}{\electronvolt} above $\varepsilon_\mathrm{F}$. (The Fermi-Dirac distribution has a smearing width $\approx 5k_\mathrm{B}T_\mathrm{e}$ at \SI{5}{\kilo\kelvin}.) For FeNi and CoNi, the present results agree with the first-principles calculations of Samolyuk \textit{et al.}~\cite{samolyukElectronPhononCoupling2016} to within approximately \SI{15}{\percent} in the ferromagnetic phase, while their non-magnetic results bracket the present PM curves (Fig.~\ref{fig:Figure_DFT}c).

Figure~\ref{fig:Figure_DFT}d shows the electrical resistivity as a function of lattice temperature, $\rho(T_\mathrm{ph})$, from the linear-response calculations (Methods, Eq.~\eqref{eq:kubo_greenwood}), including both phonon and spin disorder within the alloy-analogy model. For the ferromagnetic binary alloys, $\rho(T_\mathrm{ph})$ rises steeply with temperature as the spin disorder grows with decreasing magnetization $m(T_\mathrm{ph})$, exhibits a kink at the calculated Curie temperatures ($T_\mathrm{C} = 735$ and \SI{909}{\kelvin} for FeNi and CoNi, Methods), and saturates above $T_\mathrm{C}$. The Cr-containing alloys instead start from a large, nearly temperature-independent resistivity of \SI{70}{\micro\ohm\centi\metre} to \SI{84}{\micro\ohm\centi\metre}, characteristic of resistivity saturation in strongly disordered alloys~\cite{redkaInterplayDisorderElectronic2024}. Two computational checks isolate the role of the magnetic phase. Above $T_\mathrm{C}$ the FM curve with $m=0$ remains above the self-consistent PM/DLM result for the binary alloys, because the DLM medium collapses the local Ni moments and therefore scatters less, whereas for the ternary alloys and the quinary alloy the FM-potential ($m=0$) and PM/DLM curves nearly coincide. The calculations reproduce the experimental ordering and magnitude~\cite{redkaInterplayDisorderElectronic2024,ikedaElectricalResistivityNickel1988,samolyukTemperatureDependentElectronic2018,jinTailoringPhysicalProperties2016}, but overestimate $\rho$ for the magnetically ordered metals (\SI{29}{\micro\ohm\centi\metre} versus approximately \SI{9}{\micro\ohm\centi\metre} at \SI{300}{\kelvin} for CoNi) and underestimate it for the Cr-containing alloys (\SI{70}{\micro\ohm\centi\metre} versus \SI{110}{\micro\ohm\centi\metre} to \SI{115}{\micro\ohm\centi\metre} for CrMnFeCoNi), consistent with known limitations of the single-site CPA~\cite{redkaInterplayDisorderElectronic2024,muUncoveringElectronScattering2019}. The composition trend is most directly rationalized by the residual resistivity. Calculated and measured values from Ref.~\cite{muUncoveringElectronScattering2019} (Fig.~\ref{fig:Figure_DFT}d, hexagons at $T_\mathrm{ph}=0$) increase from below \SI{11}{\micro\ohm\centi\metre} for the binary alloys to between \SI{60}{\micro\ohm\centi\metre} and \SI{105}{\micro\ohm\centi\metre} upon Cr (and Mn) addition.

\begin{table*}[t]
  \centering
  \caption{Experimental and calculated transport and optical quantities for the investigated alloys. Calculated values refer to the room-temperature magnetic phases, FM with $m(\SI{300}{\kelvin})$ for FeNi and CoNi and PM/DLM for the Cr-containing alloys. Values marked by * are taken from Ref.~\cite{jinTailoringPhysicalProperties2016}, values marked by $^\ddagger$ from Ref.~\cite{muUncoveringElectronScattering2019}, and $^\dagger$ indicates measurements in this work.}
  \label{tab:transport-optics-comparison}
  \footnotesize
  \setlength{\tabcolsep}{5pt}
  \renewcommand{\arraystretch}{1.15}
  \sisetup{table-number-alignment=center, table-text-alignment=center, detect-weight=true, detect-family=true}
  \begin{tabular}{@{}l
      S[table-format=3.1]
      S[table-format=3.1]
      S[table-format=3.1]
      S[table-format=3.1]
    S[table-format=3.1]@{}}
    \toprule
    {Quantity} & {FeNi} & {CoNi} & {CrFeNi} & {CrCoNi} & {CrMnFeCoNi} \\
    \midrule
    \multicolumn{6}{@{}l}{\textit{Electronic thermal conductivity $\kappa_\mathrm{e}$} (\si{\watt\per\metre\per\kelvin})}\\
    Experimental (\SI{300}{\kelvin})* & 21.1 & 74.0 & 12.9 & 7.2 & 8.0 \\
    This work (calculated, \SI{300}{\kelvin}) & 12.3 & 25.0 & 9.0 & 10.3 & 10.4 \\
    \addlinespace

    \multicolumn{6}{@{}l}{\textit{Electrical resistivity $\rho$} (\si{\micro\ohm\centi\metre})}\\
    Experimental (\SI{0}{\kelvin})$^\ddagger$ & 10.7 & 2.6 & \multicolumn{1}{c}{\textemdash} & 93.5 & 101.5 \\
    KKR--CPA (\SI{0}{\kelvin})$^\ddagger$ & 3.8 & 1.4 & \multicolumn{1}{c}{\textemdash} & 67.9 & 61.2 \\
    This work (calculated, \SI{300}{\kelvin}) & 59.4 & 29.3 & 81.5 & 71.1 & 70.4 \\
    \addlinespace

    \multicolumn{6}{@{}l}{\textit{Absorptance $A$ at $\lambda=\SI{1030}{\nano\metre}$}}\\
    Measured ($\theta=\ang{0}$)$^\dagger$ & 0.306 & 0.273 & 0.275 & 0.259 & 0.301 \\
    Calculated ($\theta=\ang{0}$) & 0.266 & 0.223 & 0.249 & 0.234 & 0.237 \\
    Calculated ($\theta=\ang{36.9}$, p-pol.) & 0.320 & 0.271 & 0.301 & 0.283 & 0.287 \\
    \addlinespace

    \multicolumn{6}{@{}l}{\textit{Optical constants at $\lambda=\SI{1030}{\nano\metre}$ (calculated)}}\\
    $n - \mathrm{i}k$ & \multicolumn{1}{c}{$3.990 - 5.928\mathrm{i}$} & \multicolumn{1}{c}{$3.335 - 6.397\mathrm{i}$} & \multicolumn{1}{c}{$4.961 - 6.637\mathrm{i}$} & \multicolumn{1}{c}{$4.747 - 6.947\mathrm{i}$} & \multicolumn{1}{c}{$4.746 - 6.858\mathrm{i}$} \\
    \bottomrule
  \end{tabular}
\end{table*}

The electronic thermal conductivity, $\kappa_\mathrm{e} = L_0 T_\mathrm{ph}/\rho$ (Fig.~\ref{fig:Figure_DFT}e), follows these trends. For all alloys except the binary alloys below $T_\mathrm{C}$, $\kappa_\mathrm{e}$ increases approximately linearly with $T_\mathrm{ph}$, reaching \SI{27}{\watt\per\metre\per\kelvin} to \SI{31}{\watt\per\metre\per\kelvin} at \SI{1000}{\kelvin}. At \SI{300}{\kelvin} the calculations yield \SI{12.3}{\watt\per\metre\per\kelvin} (FeNi), \SI{25.0}{\watt\per\metre\per\kelvin} (CoNi), \SI{9.0}{\watt\per\metre\per\kelvin} (CrFeNi), \SI{10.3}{\watt\per\metre\per\kelvin} (CrCoNi), and \SI{10.4}{\watt\per\metre\per\kelvin} (CrMnFeCoNi). A direct comparison with experimental Wiedemann--Franz estimates is summarized in Table~\ref{tab:transport-optics-comparison}. The trends are robust despite numerical offsets. For the binary alloys, $\kappa_\mathrm{e}$ is highly sensitive to the magnetization curve $m(T_\mathrm{ph})$. The higher conductivity of FeNi and CoNi reflects the weakly scattered majority-spin channel of the ferromagnetic phase discussed above. By contrast, adding Cr (and further Mn) introduces strong potential and spin-disorder scattering, renders both spin channels disordered~\cite{muUncoveringElectronScattering2019}, and enhances Fermi-surface smearing~\cite{redkaInterplayDisorderElectronic2024}, which suppresses the electronic contribution to thermal conductivity through increased resistivity. At \SI{300}{\kelvin}, the phonon contribution to the total thermal conductivity is reported to lie between \SI{4}{\watt\per\metre\per\kelvin} and \SI{7}{\watt\per\metre\per\kelvin} for these alloys~\cite{jinTailoringPhysicalProperties2016}.

\begin{figure}[!ht]
  \centering
  \includegraphics[width=\linewidth]{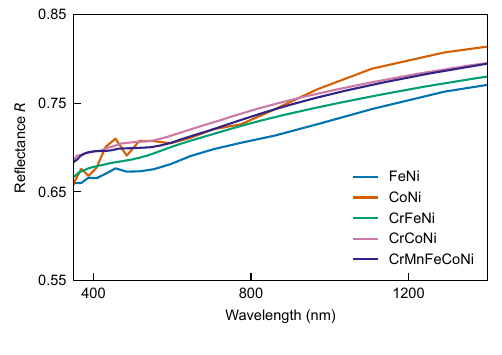}
  \caption{Calculated normal-incidence reflectance, $R$, of the Cantor--Wu alloys from the frequency-dependent Kubo--Greenwood approach (Methods, Eq.~\eqref{eq:optical_sigma1}), with FeNi and CoNi in the ferromagnetic phase with $m(T_\mathrm{ph}=\SI{300}{\kelvin})$ and the Cr-containing alloys in the paramagnetic DLM phase.}
  \label{fig:optics-R}
\end{figure}

The optical conductivity, $\sigma(\omega)$, and the complex refractive index, $n-\mathrm{i}k$, were computed from the frequency-dependent Kubo--Greenwood approach (Methods, Eq.~\eqref{eq:optical_sigma1}). The calculated reflectance at normal and non-normal incidence ($\theta=\ang{36.9}$) is only weakly composition dependent (Fig.~\ref{fig:optics-R}), in agreement with the measured spectra (Fig.~\ref{fig:methods-setup}b). All alloys show a similar, rather flat dispersion, with a stronger reflectance drop below \SI{400}{\nano\metre} and a monotonic increase towards longer wavelengths. FeNi exhibits the lowest reflectance in both experiment and theory, while the highest reflectance is found for CrCoNi in experiment and for CoNi in theory, with alloy-to-alloy differences remaining below \SI{5}{\percent}. The complex refractive index, $n - \mathrm{i}k$, and the absorptance, $A = 1 - R$, at normal incidence and at $\theta=\ang{36.9}$ (p-polarized) for \SI{1030}{\nano\metre} are listed in Table~\ref{tab:transport-optics-comparison}. The calculated absorptance at normal incidence is lower than the measured one by \SIrange{2}{5}{\percent} (absolute) across the alloys and by about \SI{6}{\percent} for CrMnFeCoNi.

\section{Discussion}
\begin{figure*}[!ht]
  \centering
  \includegraphics[width=\textwidth]{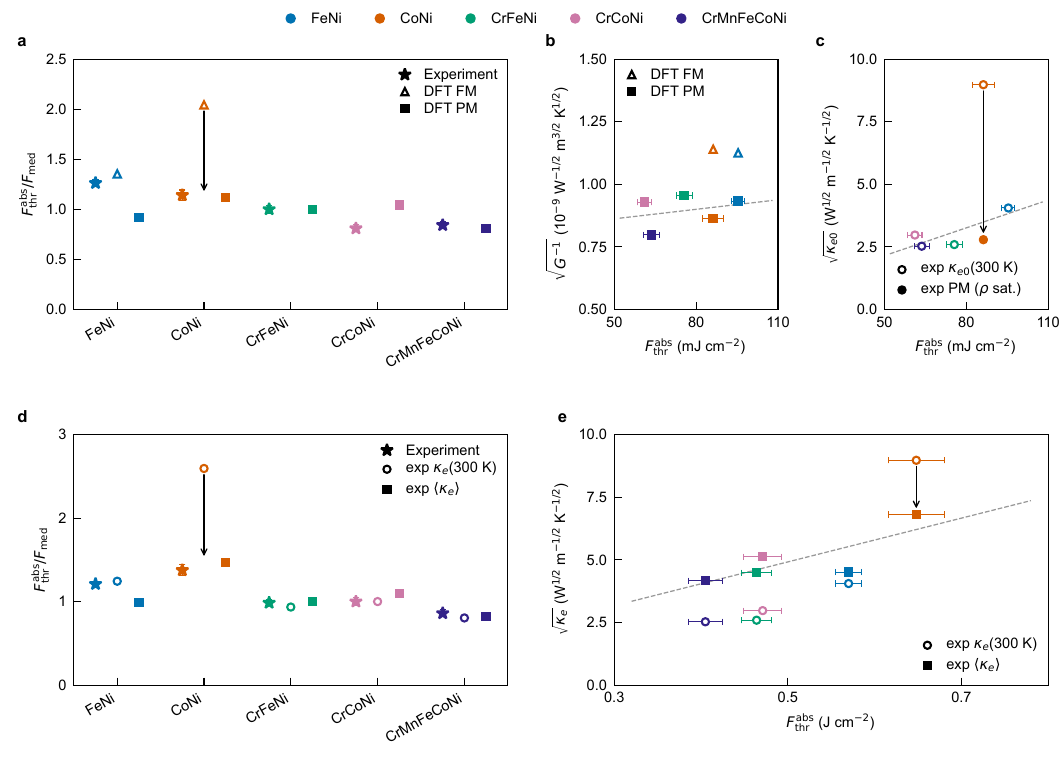}
  \caption{Transport scaling of the absorbed ablation thresholds. \textbf{a,} Measured absorbed fs ablation thresholds (stars) compared with the melting-threshold estimate of Eq.~\eqref{eq:wellershoff}, evaluated with DFT parameter sets for the ferromagnetic (FM, open triangles) and paramagnetic (PM, filled squares) phases. Experiment and each model variant are normalized to their respective medians, $F_{\mathrm{med}}$, across the alloy series. For the Cr-containing alloys the two parameter sets coincide. \textbf{b,} Inverse-root electron--phonon coupling, $\sqrt{G_\mathrm{ep}^{-1}}$, versus the measured absorbed fs ablation threshold. The dashed line is a linear guide to the eye through the PM values. \textbf{c,} Square root of the equilibrium electronic thermal conductivity $\kappa_{\mathrm{e}0}$, obtained from experimental resistivities at \SI{300}{\kelvin} via the Wiedemann--Franz law (open circles), versus the absorbed fs ablation threshold. The filled circle marks the CoNi value evaluated with the saturated high-temperature resistivity, representative of the spin-disordered phase. The dashed line is a guide to the eye through the four remaining alloys. \textbf{d,} Measured absorbed ns ablation thresholds (stars) compared with the surface-melting estimate of Eq.~\eqref{eq:ns-model}, evaluated with the experimental $\kappa_\mathrm{e}$(\SI{300}{\kelvin}) (open circles) and with the path-averaged $\langle\kappa_\mathrm{e}\rangle$ up to the melting temperature (filled squares), median-normalized as in \textbf{a}. \textbf{e,} $\sqrt{\kappa_\mathrm{e}}$ versus the absorbed ns ablation threshold for both variants, with a linear guide to the eye through the $\langle\kappa_\mathrm{e}\rangle$ values. Black arrows mark the collapse of the CoNi values upon loss of ferromagnetic order. Error bars are $1\sigma$ uncertainties of the absorbed thresholds.}
  \label{fig:discussion}
\end{figure*}

Connecting the measured ablation thresholds to the calculated transport parameters requires absorbed rather than incident fluences. For 3$d$ transition-metal alloys, including the related austenitic stainless steel AISI~304 (Fe72Cr18Ni10), the self-reflectance up to the ablation threshold during single-pulse ultrashort-pulse ablation closely matches the steady-state low-intensity reflectance~\cite{winterUltrashortSinglepulseLaser2021,pflugFluenceDependentTransientReflectance2021}. The temperature dependence of $R$ remains weak even at melting. For Ni, time-resolved measurements resolve a solid--liquid reflectivity difference of less than \SI{0.5}{\percent}~\cite{bonebergNanosecondTimeresolvedReflectivity2000}. Under ns pulse irradiation, the reflectivity of polished metals remains virtually at its room-temperature value up to the plasma-formation threshold~\cite{benavidesReflectionNanosecondNd2011}. Diffuse scattering losses are negligible, since single pulses are applied to polished surfaces. At $R\approx\SI{70}{\percent}$, the approximate reflectance of the alloys investigated here, these changes correspond to a relative change of the absorbed fluence of below approximately \SI{8}{\percent}, of the same order as the smallest differences between neighbouring alloys but well below the systematic \SIrange{36}{42}{\percent} threshold variation across the series. Because all investigated alloys show closely similar optical dispersion (Fig.~\ref{fig:methods-setup}b and Fig.~\ref{fig:optics-R}), the same approximation---using the steady-state reflectance up to the ablation threshold---applies, and the resulting trends are robust within this uncertainty. All thresholds discussed below are therefore absorbed thresholds, $F_{\mathrm{thr}}^{\mathrm{abs}}=F_{\mathrm{thr}}(1-R)$ (Table~\ref{tab:thresholds}), with $R$ taken from the measured normal-incidence reflectance at the respective laser wavelength (cf.~Methods).

A predictive, physically complete description of ablation requires, beyond the TTM, an explicit mechanistic picture of ablation, including phase transitions (ultrafast homogeneous melting), vaporization or phase explosion, isochoric pressure build-up, and mechanical unloading that can drive tensile failure and spallation~\cite{rethfeldModellingUltrafastLaser2017,zhigileiAtomisticModelingShort2009}. Such mechanisms are commonly implemented via hydrodynamic~\cite{povarnitsynMultimaterialTwotemperatureModel2007} or molecular-dynamics~\cite{ivanovCombinedAtomisticcontinuumModeling2003} extensions of the TTM and therefore demand reliable interatomic potentials or multiphase equations of state, a requirement that is particularly restrictive for compositionally complex alloys. A practical alternative is to relate the measured ablation thresholds to a reduced set of transport parameters within TTM-based scaling. Following Corkum \textit{et al.}~\cite{corkumThermalResponseMetals1988} and Wellershoff \textit{et al.}~\cite{wellershoffRoleElectronPhonon1999}, the electronic thermal conductivity remains enhanced during electron--phonon equilibration, $\kappa_\mathrm{e}\approx\kappa_{\mathrm{e}0}\,T_\mathrm{e}/T_\mathrm{ph}$, and with $C_\mathrm{e}\approx\gamma T_\mathrm{e}$ and a constant coupling factor $G_\mathrm{ep}$ an analytic estimate of the melting-threshold fluence is obtained,
\begin{equation}
  F_{\mathrm{thr}}^{\mathrm{melt}} = \left(\frac{128}{\pi}\right)^{1/8}\left(\frac{\kappa_{\mathrm{e}0}^2\, C_{\mathrm{ph}}^5\, T_\mathrm{m}^3}{\gamma\, G_\mathrm{ep}^2}\right)^{1/4},
  \label{eq:wellershoff}
\end{equation}
where $\kappa_{\mathrm{e}0}$ is the equilibrium electronic thermal conductivity, $C_{\mathrm{ph}}$ the lattice heat capacity, and $T_\mathrm{m}$ the melting temperature. Equation~\eqref{eq:wellershoff} demonstrates why reduced electronic heat conduction and larger electron--phonon coupling lower characteristic ablation thresholds. Ballistic transport corrections can be neglected here, as the electron mean free paths in concentrated solid solutions are on the order of a nanometre~\cite{robartsExtremeFermiSurface2020}, well below the measured optical penetration depths of \SI{15}{\nano\metre} to \SI{16}{\nano\metre}~\cite{redkaSubpicosecondSinglepulseLaser2021}.

We evaluate Eq.~\eqref{eq:wellershoff} with all inputs taken consistently from one magnetic phase. For FeNi and CoNi, the FM parameter set uses the ferromagnetic results of Fig.~\ref{fig:Figure_DFT}. The equilibrium conductivities $\kappa_{\mathrm{e}0}$ at \SI{300}{\kelvin} are \SI{12.3}{\watt\per\metre\per\kelvin} and \SI{25.0}{\watt\per\metre\per\kelvin}, the Sommerfeld coefficients $\gamma$ are \SI{481}{\joule\per\metre\cubed\per\kelvin\squared} and \SI{520}{\joule\per\metre\cubed\per\kelvin\squared}, and the low-$T_\mathrm{e}$ coupling factors $G_\mathrm{ep}$ are \SI{0.79e18}{\watt\per\metre\cubed\per\kelvin} and \SI{0.77e18}{\watt\per\metre\cubed\per\kelvin}. The PM set uses the corresponding DLM results. Here $\kappa_{\mathrm{e}0}$ is \SI{8.1}{\watt\per\metre\per\kelvin} and \SI{13.2}{\watt\per\metre\per\kelvin}, $\gamma$ is \SI{559}{\joule\per\metre\cubed\per\kelvin\squared} and \SI{629}{\joule\per\metre\cubed\per\kelvin\squared}, and $G_\mathrm{ep}$ is \SI{1.15e18}{\watt\per\metre\cubed\per\kelvin} and \SI{1.34e18}{\watt\per\metre\cubed\per\kelvin}. For the Cr-containing alloys, already paramagnetic at room temperature, only the PM/DLM parameter set applies. The lattice heat capacity is taken in the Dulong--Petit limit, $C_{\mathrm{ph}}=12k_\mathrm{B}/a_{\mathrm{fcc}}^3$, giving \SI{3.55}{\mega\joule\per\metre\cubed\per\kelvin} to \SI{3.75}{\mega\joule\per\metre\cubed\per\kelvin}, and the melting temperatures are \SI{1703}{\kelvin}, \SI{1735}{\kelvin}, \SI{1690}{\kelvin}, and \SI{1553}{\kelvin} for FeNi, CoNi, CrCoNi, and CrMnFeCoNi, measured by differential scanning calorimetry~\cite{wuTemperatureDependenceMechanical2014}. For CrFeNi the value of \SI{1730}{\kelvin} of the closely related stainless steel AISI~304 is adopted~\cite{redkaSubpicosecondSinglepulseLaser2021}.
The absolute estimates of Eq.~\eqref{eq:wellershoff} lie substantially below the measured absorbed ablation thresholds, as expected for a melting-onset criterion that neglects the ablation energetics (latent heats and overheating). The comparison in Fig.~\ref{fig:discussion}a is therefore made on median-normalized values, $F/F_{\mathrm{med}}$, where $F_{\mathrm{med}}$ is the median absorbed ablation threshold across the five alloys, taken separately for the experiment and for each model variant, which isolates the trends across compositions from the global offset.

Figure~\ref{fig:discussion}a shows that the PM parameter set reproduces the measured hierarchy across the entire series, including the decrease upon Cr addition, with a maximum deviation of \SI{28}{\percent} for FeNi and near-quantitative agreement for CoNi. The FM set instead overestimates the CoNi threshold by approximately \SI{80}{\percent} (arrow in Fig.~\ref{fig:discussion}a), which can be explained as follows. In the ferromagnetic phase at \SI{300}{\kelvin}, spin-disorder scattering is largely frozen out and CoNi in particular, with its high Curie temperature (measured \SI{1117}{\kelvin}~\cite{jinThermophysicalPropertiesNicontaining2017}, calculated \SI{909}{\kelvin}, Methods), retains a nearly pristine band structure in the minority-spin channel~\cite{muUncoveringElectronScattering2019} and hence a large $\kappa_{\mathrm{e}0}$ together with a small $G_\mathrm{ep}$ (results). Under fs excitation, however, this ordered phase does not survive the energy deposition. Ultrafast demagnetization proceeds within a few hundred femtoseconds in 3$d$ ferromagnets~\cite{beaurepaireUltrafastSpinDynamics1996a,koopmansExplainingParadoxicalDiversity2010,eichBandStructureEvolution2017}, faster than the electron--phonon equilibration time ($\tau_\mathrm{ep}\approx\SIrange{1}{2}{\pico\second}$) that sets the transport-relevant window in Eq.~\eqref{eq:wellershoff}. This transition is not a Stoner-type collapse of the exchange splitting~\cite{stonerCollectiveElectronFerromagnetism1938}. Instead, local moments persist and lose their long-range alignment through transverse fluctuations~\cite{eichBandStructureEvolution2017}, which is precisely the spin configuration described by the DLM medium underlying the PM calculations. The fs ablation threshold therefore probes the transport parameters of the spin-disordered phase. For FeNi, the FM and PM sets differ much less, and for the Cr-containing alloys, paramagnetic already at room temperature, the distinction is absent altogether.

The scaling structure of Eq.~\eqref{eq:wellershoff} follows from a minimal energy balance, $F_{\mathrm{thr}}\propto d_{\mathrm{diff}}\approx\sqrt{D_\mathrm{e}\tau_\mathrm{ep}}$, with $D_\mathrm{e}=\kappa_\mathrm{e}/C_\mathrm{e}$ and $\tau_\mathrm{ep}=C_\mathrm{e}/G_\mathrm{ep}$, yielding $F_{\mathrm{thr}}\propto\sqrt{\kappa_\mathrm{e}/G_\mathrm{ep}}$~\cite{wellershoffRoleElectronPhonon1999,winterUltrashortSinglepulseLaser2021}. Figures~\ref{fig:discussion}b,c decompose the measured absorbed fs ablation thresholds along these two axes. In Fig.~\ref{fig:discussion}b, $\sqrt{G_\mathrm{ep}^{-1}}$ of the PM phase correlates with the absorbed fs threshold across all five alloys, following the linear guide to the eye, while the FM values of the binary alloys lie visibly above.

Figure~\ref{fig:discussion}c examines the dependency of the ablation threshold on the conductivity, using experimental rather than calculated $\kappa_{\mathrm{e}0}$, obtained from the measured resistivities via the Wiedemann--Franz law, since the single-site CPA systematically compresses the conductivity contrast across compositions (Sec.~\ref{sec:results-dft}). At \SI{300}{\kelvin} the four alloys FeNi, CrFeNi, CrCoNi, and CrMnFeCoNi follow a near-linear $\sqrt{\kappa_{\mathrm{e}0}}$ correlation with the fs threshold, whereas CoNi, with $\kappa_{\mathrm{e}0}=\SI{81}{\watt\per\metre\per\kelvin}$, lies an order of magnitude above this band. Evaluating $\kappa_{\mathrm{e}0}$ instead with the resistivity measured above the Curie temperature, where the spin-disorder contribution has saturated~\cite{ikedaElectricalResistivityNickel1988}, collapses the CoNi value to \SI{7.8}{\watt\per\metre\per\kelvin} (arrow in Fig.~\ref{fig:discussion}c), directly into the band of the remaining alloys.

The nanosecond ablation thresholds provide an independent picture of the conductivities, in this case probed in thermal equilibrium. For $\tau_\mathrm{p}=\SI{102}{\nano\second}\gg\tau_\mathrm{ep}$, a one-temperature Fourier description applies~\cite{corkumThermalResponseMetals1988,stuartNanosecondtofemtosecondLaserinducedBreakdown1996}, and the standard solution for one-dimensional heat flow into a semi-infinite solid at constant absorbed flux places the onset of surface melting at
\begin{equation}
  F_{\mathrm{thr}}^{\mathrm{melt}} = \frac{T_\mathrm{m} - T_0}{2}\sqrt{\pi\,\kappa\, C_{\mathrm{ph}}\,\tau_\mathrm{p}},
  \label{eq:ns-model}
\end{equation}
with $T_0=\SI{300}{\kelvin}$ and the total conductivity $\kappa=\kappa_\mathrm{e}+\kappa_{\mathrm{ph}}$, where $\kappa_{\mathrm{ph}}\approx\SI{5}{\watt\per\metre\per\kelvin}$ is adopted for all alloys, representative of the reported range~\cite{jinTailoringPhysicalProperties2016}. In contrast to the fs pulse duration, the material is now heated in quasi-equilibrium from \SI{300}{\kelvin} to $T_\mathrm{m}$, and the relevant conductivity is the average along the heating path. We therefore evaluate Eq.~\eqref{eq:ns-model} with two experimental input sets, the \SI{300}{\kelvin} value $\kappa_\mathrm{e}(\SI{300}{\kelvin})$ and the path average
\begin{equation}
  \langle\kappa_\mathrm{e}\rangle=\frac{1}{T_\mathrm{m}-T_0}\int_{T_0}^{T_\mathrm{m}} \frac{L_0\, T}{\rho(T)}\,\mathrm{d}T,
  \label{eq:kappa-path-average}
\end{equation}
with $\rho(T)$ interpolated from measured resistivities~\cite{redkaInterplayDisorderElectronic2024,ikedaElectricalResistivityNickel1988,samolyukTemperatureDependentElectronic2018} and held at its boundary values outside the measured range. The absolute estimates fall only approximately \SI{20}{\percent} below the measured absorbed thresholds, remarkably close for a melting-onset criterion, and Fig.~\ref{fig:discussion}d again compares median-normalized values. With $\kappa_\mathrm{e}(\SI{300}{\kelvin})$, CoNi is overestimated by approximately \SI{90}{\percent}, for the same reason as for the fs pulse duration. With $\langle\kappa_\mathrm{e}\rangle$, which averages across the ferromagnetic-to-paramagnetic transition, the CoNi prediction drops to 1.48 (arrow in Fig.~\ref{fig:discussion}d), close to the measured 1.34, and the full series is reproduced within \SI{17}{\percent}. The corresponding $\sqrt{\kappa_\mathrm{e}}$ correlation (Fig.~\ref{fig:discussion}e) shows the same collapse, with the path-averaged values of all alloys following the linear guide to the eye. CoNi retains the highest threshold for the ns pulse duration, in contrast to the fs case, because its conductivity advantage survives in the path average, whereas under fs excitation the demagnetization quenches it before the energy is transferred to the lattice.

\section{Conclusion}
Femtosecond and nanosecond single-pulse ablation thresholds of the five selected Cantor--Wu alloys of increasing compositional complexity were measured under identical conditions, corrected for absorption using reflectance spectra measured on the same surfaces, and interpreted through first-principles-derived transport parameters obtained using the KKR--CPA. Although the absorbed threshold fluences vary by less than a factor of 1.6 within each pulse duration, the two pulse durations track complementary transport properties. The femtosecond thresholds probe the electron--phonon coupling and the electronic conductivity of the transiently spin-disordered phase, reproduced across the full series by a two-temperature-model scaling with all inputs taken from the paramagnetic DLM calculations. The nanosecond thresholds probe the equilibrium conductivity averaged along the heating path to melting, captured by a one-temperature surface-melting estimate to within approximately \SI{20}{\percent} in absolute terms.

The behaviour of CoNi demonstrates how the thresholds can sense the magnetic phase. Its room-temperature transport parameters, dominated by the weakly scattered majority-spin channel of the ferromagnetic phase, overpredict the thresholds by up to a factor of two for both pulse durations. Evaluating the same models with the spin-disordered conductivity, reached transiently through ultrafast demagnetization under femtosecond excitation and thermally by heating across the Curie temperature in the nanosecond case, restores quantitative agreement. The ablation threshold of a magnetic alloy is thus set not by its tabulated room-temperature transport but by the transport properties of the phase created by the laser pulse itself.

These results establish single-pulse ablation thresholds, an experimentally simple and contact-free observable, as quantitative probes of electronic transport and its magnetic-phase dependence in compositionally complex alloys. Conversely, they show that predictive threshold estimates for alloy families are within reach of purely \textit{ab initio} parameter sets for compositionally complex alloys and their magnetic phases.

\section*{Author contributions}
D.R.: conceptualization, methodology, formal analysis, investigation, data curation, visualization, writing -- original draft. M.S.: investigation, supervision, writing -- review \& editing. R.B.: investigation. C.D.W.: investigation, writing -- review \& editing. H.E.: software, writing -- review \& editing. J.M.: software, funding acquisition, writing -- review \& editing. D.J.F.: supervision, writing -- review \& editing. H.P.H.: funding acquisition, writing -- review \& editing.

\section*{Data availability}
The data that support the findings of this study, including the raw and processed datasets underlying the figures, are available from the corresponding author upon reasonable request.

\begin{acknowledgments}
The authors thank Kilian Sandner and Uwe Glatzel (Metals and Alloys, University of Bayreuth) for their support with the alloy synthesis, and Constanze Eulenkamp for the SEM imaging. This work was supported by the Deutsche Forschungsgemeinschaft (Grant No.~528706678) and by the project MEBIOSYS (Project No.~CZ.02.01.01/00/22\_008/0004634) funded by the Programme Johannes Amos Commenius (call Excellent Research). C.D.W. acknowledges support from a UK Engineering and Physical Sciences Research Council (EPSRC) Doctoral Prize Fellowship at the University of Bristol, Grant EP/W524414/1. The authors also acknowledge the use of compute resources provided by the Isambard 3 high-performance computing (HPC) facility. Isambard 3 is hosted by the University of Bristol and operated by the GW4 Alliance (\href{https://gw4.ac.uk}{https://gw4.ac.uk}) and is funded by UK Research and Innovation; and the EPSRC, Grant EP/X039137/1. The authors declare no competing interests.
\end{acknowledgments}

\bibliography{references}

\end{document}